\documentclass[prd, twocolumn, amsmath, superscriptaddress, floatfix, nofootinbib]{revtex4-1}

\usepackage{graphicx,amssymb,amsmath,amsthm,amsfonts}
\usepackage[linktocpage]{hyperref}

\usepackage{xcolor}
\usepackage{bm}
\usepackage{comment}
\usepackage{nicefrac}

\begin{document}

\title{Self-interacting dipolar boson stars and their dynamics}

\author{Pedro~Ildefonso}
\affiliation{Institute for Theoretical Physics, University of Innsbruck, A-6020 Innsbruck, Austria}
\affiliation{Departamento  de  Matem\'{a}tica  da  Universidade  de  Aveiro and 
  Centre for Research and Development in Mathematics and Applications (CIDMA),
  Campus de Santiago, 3810-183 Aveiro, Portugal}
\affiliation{Centro de Astrof\'\i sica e Gravita\c c\~ao -- CENTRA,
  Departamento de F\'\i sica, Instituto Superior T\'ecnico -- IST,
  Universidade de Lisboa -- UL, Avenida Rovisco Pais 1, 1049-001 Lisboa, Portugal}

\author{Miguel~Zilh\~ao}
\affiliation{Departamento  de  Matem\'{a}tica  da  Universidade  de  Aveiro and 
  Centre for Research and Development in Mathematics and Applications (CIDMA),
  Campus de Santiago, 3810-183 Aveiro, Portugal}

\author{Carlos~Herdeiro}
\affiliation{Departamento  de  Matem\'{a}tica  da  Universidade  de  Aveiro and 
  Centre for Research and Development in Mathematics and Applications (CIDMA),
  Campus de Santiago, 3810-183 Aveiro, Portugal}

\author{Eugen~Radu}
\affiliation{Departamento  de  Matem\'{a}tica  da  Universidade  de  Aveiro and 
  Centre for Research and Development in Mathematics and Applications (CIDMA),
  Campus de Santiago, 3810-183 Aveiro, Portugal}

\author{Nuno M. Santos}
\affiliation{Departamento  de  Matem\'{a}tica  da  Universidade  de  Aveiro and 
  Centre for Research and Development in Mathematics and Applications (CIDMA),
  Campus de Santiago, 3810-183 Aveiro, Portugal}
\affiliation{Centro de Astrof\'\i sica e Gravita\c c\~ao -- CENTRA,
  Departamento de F\'\i sica, Instituto Superior T\'ecnico -- IST,
  Universidade de Lisboa -- UL, Avenida Rovisco Pais 1, 1049-001 Lisboa, Portugal}

\begin{abstract}
We construct and dynamically evolve dipolar, self-interacting scalar boson stars in a model with sextic (+ quartic) self-interactions. The domain of existence of such dipolar \textit{$Q$-stars} has a similar structure to that of the fundamental monopolar stars of the same model. For the latter it is structured in a Newtonian plus a relativistic branch, wherein perturbatively stable solutions exist, connected by a middle unstable branch. Our evolutions support similar dynamical properties of the dipolar $Q$-stars that: 1)  in the Newtonian and relativistic branches are dynamically robust over time scales longer than those for which dipolar stars without self-interactions are seen to decay; 2) in the middle branch migrate to either the Newtonian or the relativistic branch; 3) beyond the relativistic branch decay to black holes. Overall, these results strengthen the observation, seen in other contexts, that self-interactions can mitigate dynamical instabilities of scalar boson star models.
\end{abstract}

\maketitle

\section{Introduction}
\label{sec:intro}

As it is by now well-understood, Einstein's gravity minimally coupled to massive scalar fields gives rise to macroscopic stable configurations named \textit{boson stars} (BSs) \cite{PhysRev.172.1331, PhysRevD.77.044036,PhysRev.187.1767, doi:10.1063/1.1703887, PhysRev.168.1445, PhysRevD.12.319} -- see~\cite{mielke2003topical,Shnir:2022lba} for reviews.
This class of compact objects comprises a large group of different models, many of which proven to be dynamically robust -- see~\cite{Liebling:2012fv} for a review and~\cite{Seidel:1990jh,PhysRevD.75.064005,PhysRevD.96.104040,PhysRevD.95.124005,Sanchis-Gual:2017bhw,PhysRevLett.126.241105,Sanchis-Gual:2019ljs,DiGiovanni:2020ror,Evstafyeva:2022bpr,Jaramillo:2022zwg,Sanchis-Gual:2021phr,Brito:2023fwr,Bezares:2022obu} for specific dynamical analyses, also for the case of the cousin \textit{vector} BS ($aka$ Proca) model. 

Among the models of BSs are those comprising a scalar potential free of self-interactions, namely ``mini-BSs''~\cite{LEE1989477}, and those possessing self-interactions, such as ``Q-stars'' \cite{doi:10.1063/1.1664693, lee1974vacuum, lee1976symposium, PhysRevD.13.2739, COLEMAN1985263, PhysRevD.35.3658, LYNN1989465, LEE1992251,Boskovic:2021nfs}. Due to their dynamical robustness, a class of those have shown to be good black hole~(BH) mimickers, in the sense, for instance, of being able to match the predictions made for the merger of two BHs and used to interpret real gravitational-wave signals~\cite{PhysRevLett.126.081101,CalderonBustillo:2022cja}, as well as mimicking the (effective) shadow of a BH~\cite{Olivares:2018abq,Herdeiro_2021,Rosa:2022tfv}. Their role as BHs mimickers in a variety of models \cite{Cardoso2019}, and their appeal as candidates for some of the dark matter in our Universe \cite{sharma2008boson}, in particular within the fuzzy dark matter paradigm~\cite{Hui:2016ltb,freitas2021ultralight}, support their astrophysical interest. Moreover, recent advances in gravitational-wave astronomy, \textit{e.g.}~the increasing precision of gravitational wave detectors \cite{Danzmann_1996, bender1998lisa, PhysRevD.72.083005}, place us on the verge of discovering new and more accurate results capable of distinguishing the nature and behavior of these compact objects, which has led to the effort of building up the first waveform catalog of signals sourced by exotic compact objects, namely (vector) BSs~\cite{Sanchis-Gual:2022mkk}.

Establishing the dynamical robustness of different models of BSs forms an essential theoretical basis for their possible occurrence in nature and therefore for their use in the analysis of experimental data. In this respect, it has been recently observed that scalar field self-interactions can mitigate the instability, or quench it altogether, of some excited BSs solutions, namely rotating~\cite{DiGiovanni:2020ror,Siemonsen:2020hcg}, or radially excited~\cite{Sanchis-Gual:2021phr,Brito:2023fwr}. It is therefore natural to ask whether a similar strengthening of dynamical robustness can be observed in other models of excited BSs by virtue of self-interactions. 

A less explored model of excited BSs, in particular concerning their dynamics, is the model of \textit{multipolar} BSs~\cite{HERDEIRO2021136027}. These are static (non-rotating) BSs but which have a multipolar morphology in their energy distribution, like hydrogen orbitals have a multipolar distribution for their probability density, with the spherical orbitals being a mere special case -- the $Ns$  orbitals, $N\in \mathbb{N}$. Similarly, within the multipolar family, spherical BSs are a mere special case, containing both the very fundamental stars and also the radially excited states. The simplest non-spherical multipolar BSs are the \textit{dipolar} ones~\cite{Cunha:2022tvk}, akin to $p$-orbitals.  These are two-center solitons, with a $\mathbb{Z}_2$-even metric, defining an equatorial plane above/below which a scalar lump is found, but with a  $\mathbb{Z}_2$-odd scalar field -- hence a dipole. They can also be interpreted as two monopolar BSs in equilibrium, with their gravitational attraction balanced by their scalar repulsion, as a result of the $\pi$ phase difference between the north and south hemispheres~\cite{PhysRevD.75.064005}. Dipolar BSs can also be made to spin and, in that case, be in equilibrium with one~\cite{Kunz:2019bhm} or two~\cite{Herdeiro:2023roz} (also balanced) spinning BHs.  

A study of the stability of dipolar BSs was reported in~\cite{PhysRevLett.126.241105}, wherein the (few) cases studied were shown to decay to the spherical fundamental stars. Here, we further explore the dynamical stability of dipolar stars, via non-linear dynamical evolutions, focusing on the effect of adding self-interactions. Specifically, we construct \textit{dipolar $Q$-stars} in a model with sextic (+ quartic) self-interactions. We show their domain of existence resembles that seen for the monopolar stars of the same model. Moreover, we provide evidence from our numerical evolutions that the self-interactions can increase the dynamical robustness of the dipolar stars, as in the case of rotating BSs and radially excited spherical BSs, and that the stability of the dipolar solutions bears a resemblance with that observed for the perturbative stability of monopolar stars of the same model.

This paper is organized as follows. In Section \ref{sec:backg}, we discuss equilibrium BSs, reviewing both fundamental spherical and excited dipolar BSs. As a novel result, we construct dipolar $Q$-stars with sextic (+ quartic) self-interactions, briefly discussing their main properties and also discussing the stability of the monopolar stars in the same model. In Section \ref{sec:frame}, we cover the mathematical formalism and the computational framework with which we performed the numerical simulations.
We show and discuss our results for the evolutions in Section \ref{sec:Dipolar Boson Star solutions}, where we evaluate the dynamical robustness of the dipolar $Q$-stars. We close with a discussion and comments in Section~\ref{sec:concl}. We use natural units $c=G=1$ throughout.

\section{Dipolar $Q$-stars}
\label{sec:backg}
The action $S$ for Einstein's gravity minimally coupled to a complex (massive) scalar field $\phi$ reads
\begin{align}
    S & = \int \text{d}^{4} x \sqrt{-g}\ \left[ \frac{R}{16 \pi}
    - g^{a b} \partial_a \phi^{*} \partial_b \phi - U \left(|\phi|^2 \right)  \right]\ .
    \label{eq:action}
\end{align}
The corresponding equations of motion are
\begin{align}
    &R_{ab} - \frac{1}{2} g_{ab} R = 8 \pi T_{ab}\ , \label{eq:efes_natural} \\
    &\square \phi = \frac{\partial U}{\partial |\phi|^2} \phi\ , \label{eq:kg}
\end{align}
where the stress-energy tensor reads
\begin{align}
    T_{ab} & =  \nabla_a \phi^* \nabla_b \phi + \nabla_b \phi^* \nabla_a \phi \nonumber \\
    & \quad - g_{ab} \left[ \nabla_c \phi^* \nabla^c \phi + U\left(|\phi|^2\right) \right]\ ,
    \label{eq:Tab}
\end{align}
and $\square\equiv\nabla^a \nabla_a$. 

The action (\ref{eq:action}) is invariant under the {\it global} \, $U(1)$ transformation $\phi\rightarrow e^{i\alpha}\phi$, where $\alpha$ is a constant, 
which implies the existence  of a conserved current, $j^a=-i (\phi^* \partial^a \phi-\phi \partial^a \phi^*)$, with  $\nabla_a j^a=0$.
Therefore, integrating the timelike component of this 4-current on a spacelike slice $\Sigma$ 
results in a conserved quantity -- the \textit{Noether charge}:
\begin{eqnarray}
\label{Q}
Q=\int_{\Sigma}~j^t\ ,
\end{eqnarray}
which corresponds to the number of scalar particles (upon quantization).  

Together, equations \eqref{eq:efes_natural} and \eqref{eq:kg} compose the Einstein-Klein-Gordon (EKG) system of equations. One family of solutions of these equations are self-gravitating solitons, or BSs, of which we now discuss specific members.

\begin{figure}[h!]
    \centering
    \includegraphics[width = 0.49\textwidth]{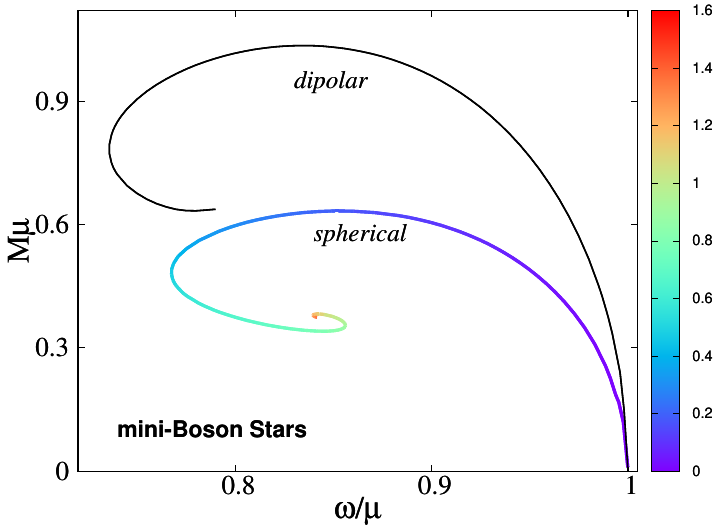} \includegraphics[width = 0.48\textwidth]{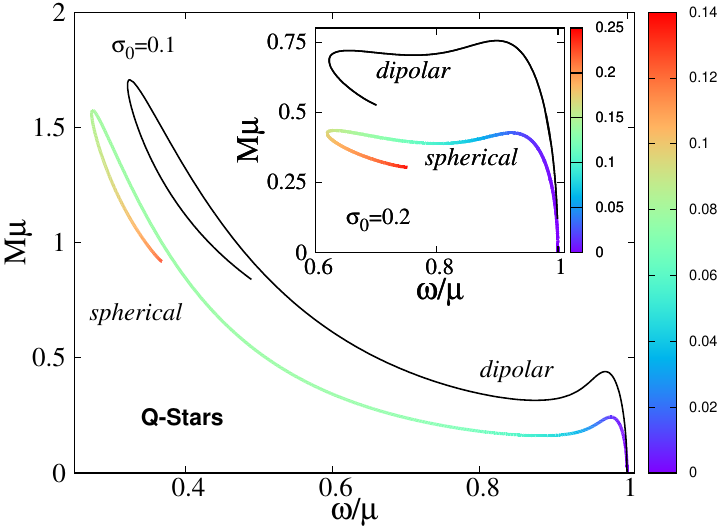} 
    \caption{\emph{Mini-BSs (top panel) and $Q$-stars (bottom panel, for $\sigma_0=0.1, 0.2$) for both spherical and dipolar BSs.} Spherical mini-BSs are perturbatively stable between the maximal frequency $\omega/\mu=1$ and the maximal mass at $(\omega/\mu, M\mu, \phi_0(0))=(0.853, 0.633, 0.192)$. Spherical $Q$-stars are perturbatively stable between the maximal frequency $\omega/\mu=1$ and the local maximum of the mass at (for $\sigma_0=0.2$) $(\omega/\mu, M\mu, \phi_0(0))=(0.923, 0.426, 0.031)$ (Newtonian stable branch) and between the local minimum of the mass at $(\omega/\mu, M\mu, \phi_0(0))=(0.802, 0.388, 0.096)$ and the global maximum of the mass at $(\omega/\mu, M\mu, \phi_0(0))=(0.63, 0.435, 0.159)$ (relativistic stable branch). The colour bar gives $\phi_0(0)$ for spherical BSs.
    \label{fig1}}
\end{figure}

BSs in the free-field  model,
\begin{equation}
    U\left(|\phi|^2\right) = \mu^2 |\phi|^2\ ,
    \label{eq:free_U}
\end{equation}
are known as \textit{mini-BSs}. Their fundamental states correspond to 
 (nodeless) spherically-symmetric scalar field distributions,
\begin{equation}
    \phi\left(r, t\right) = \phi_0 \left(r\right) e^{-i \omega t}\ ,
    \label{eq:scalar_field_sphe}
\end{equation}
where $\omega$ is the oscillation frequency and $\phi_0\left(r\right) \sim e^{-r \sqrt{\mu^2 - \omega^2}}/r$ the real, radially asymptotic profile function. These possess an established formation mechanism \cite{PhysRevLett.72.2516} while fulfilling also the criteria of dynamical stability \cite{GLEISER1989733, LEE1989477} in one branch of the domain of existence, that connects the Newtonian limit $\omega/\mu \to 1$ to the maximal mass solution -- see Fig.~\ref{fig1} (top panel). 

\begin{figure}[h!]
    \centering
    \includegraphics[width=0.95\columnwidth]{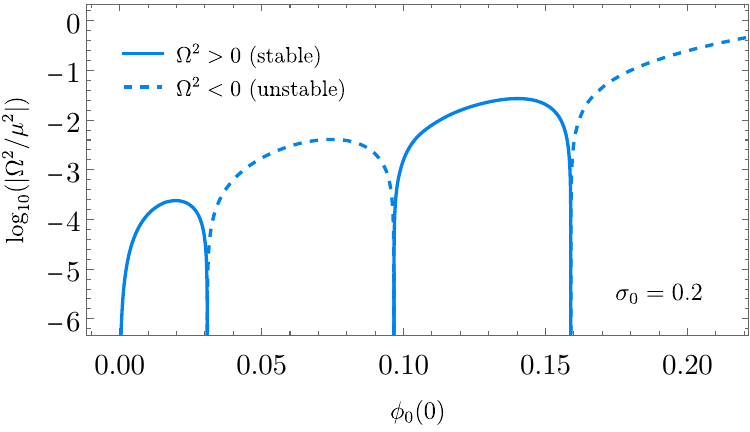} 
    \caption{\emph{Spherical perturbations with frequency $\Omega$ of the spherical $Q$-stars in~Fig.~\ref{fig1} (bottom, inset).} $\Omega^2$ changes sign precisely at the extremes of the mass. Solid (dashed) curves correspond to $\Omega^2>0$ ($\Omega^2<0$), wherein the stars are perturbatively stable (unstable) against such spherical perturbations. The setup and a detailed description of these perturbations will be presented elsewhere~\cite{SantosTA}.
    \label{fig2}}
\end{figure}

On the other hand, BSs whose scalar field obeys a sextic (+ quartic) self-interacting scalar potential are dubbed \textit{Q-stars}, since for this potential there are flat spacetime solutions called \textit{Q-balls}~\cite{COLEMAN1985263}. Here we shall consider a specific model within this sextic class of potentials, namely
\begin{align}
    U\left(|\phi|^2\right) &= \mu^2\ |\phi|^2 \bigg[ 1 -  \frac{2}{\sigma_0^2} |\phi|^2 \bigg]^2 \label{eq:qball_U} 
\end{align}
where the parameter $\sigma_0^2$ determines the compactness of the star. In this model, $Q$-stars  may become very compact and with an almost step-function decay of the scalar field and the energy density -- \textit{cf.}~Fig.~\ref{fig:1sol_ex}. Generically, spherical, fundamental $Q$-stars also possess a known formation \cite{AFFLECK1985361} and stability \cite{KUSMARTSEV199224, PhysRevD.43.3895} mechanisms. Their domain of existence is now more involved -- Fig.~\ref{fig1} (bottom panel). As $\sigma_0$ decreases and self-interactions become stronger, the spiral shape seen in Fig.~\ref{fig1} (top panel) shifts into the ``duck-like" curve seen in Fig.~\ref{fig1} (bottom panel, inset and main), possessing 3 extrema before the minimum frequency is attained. Then, there are two disconnected stable branches within a perturbative analysis: a \textit{Newtonian} stable branch, connecting the maximum allowed frequency to the first maximum of the ADM mass; and a \textit{relativistic} stable branch, connecting a local minimum of the mass to the second (global, for the plotted $\sigma_0$) maximum of the mass.
In between these branches, one finds a \textit{middle} unstable branch, and beyond (for smaller frequencies) the relativistic branch one finds (at least) another branch of very compact unstable solutions.
In Fig.~\ref{fig2} we show the result of the corresponding perturbative analysis, establishing the above conclusion for $\sigma_0=0.2$. 

In both models above, there are are also \emph{excited BSs}, besides the spherical fundamental ones, which occur in various guises. Here, we are interested in the static, non-spherical sector, \textit{i.e.}~multipolar BSs, introduced in~\cite{HERDEIRO2021136027} in the model without self-interactions. We shall focus our attention on the dipolar stars -- see Fig.~\ref{fig1} (top and bottom panels) for the domain existence of dipolar mini-BS and dipolar $Q$-stars, compared to the one of the spherical stars in the same model\footnote{Axisymmetric ``chains" (with more than two centres) of BSs have also been considered in the literature, with~\cite{Herdeiro:2021mol,Gervalle:2022fze} or without self-interactions~\cite{Sun:2022duv}.} (see~\cite{Yoshida:1997nd} for an early discussion of dipolar BSs). Dipolar stars are described by an axisymmetric scalar field
\begin{align}
    \phi\left(t,r,\theta \right) = \phi_0\left(r,\theta\right) e^{- i \omega t}\ ,
    \label{eq:scalar_field_axia}
\end{align}
which is odd parity, \textit{i.e.}~$\phi_0\left(r,\theta\right)=-\phi_0\left(r,\pi-\theta\right)$.
To construct the odd-parity static BSs with the potential~(\ref{eq:qball_U}), which were not discussed previously in the literature, the dipolar $Q$-stars,
we use a line-element with
two commuting Killing vector fields,
$\xi$ and $\eta$, with
$ \xi=\partial_t$
$\eta=\partial_{\varphi} $
in a system of adapted coordinates.
We consider the generic axisymmetric ansatz 
\begin{align}
    \text{d}s^2 = - e^{2 F_0\left(r, \theta\right)}\text{d}t^2 &
    + e^{2 F_1\left(r, \theta\right)} \left(\text{d}r^2 + r^2\text{d}\theta^2 \right) \nonumber \\ 
    &+ e^{2 F_2\left(r, \theta\right)} r^2 \sin^2{\theta}\text{d}\varphi^2\ ,
\label{eq:g_isot_axi_num}
\end{align}
in terms of the three metric functions $F_{0,1,2}$. 
The equilibrium dipolar solutions are constructed by solving numerically the EKG equations, following~\cite{Cunha:2022tvk} -- see also  \cite{HERDEIRO2015} for details -- with specified boundary conditions that we now describe. 

At the origin, spatial infinity and on the axis, the metric functions and the scalar field profile obey
\begin{align*}
	\partial_r F_{0,1,2}\rvert_{r=0} = \partial_r \phi_0\rvert_{r=0} &= 0\ , \\
	F_{0,1,2}\rvert_{r=\infty} = \phi_0\rvert_{r=\infty} &= 0\ ,\\
	\partial_{\theta} F_{0,1,2}\rvert_{\theta=0, \pi} = \partial_{\theta} \phi_0\rvert_{\theta=0, \pi} &= 0\ .
\end{align*}
Additionally, in accordance to the parity discussed above, the metric functions are invariant \textit{w.r.t.}\ a reflection along the equatorial plane, $\theta = \pi/2$, while the scalar field changes sign. This implies the equatorial boundary conditions
\begin{equation*}
	\partial_{\theta} F_{0,1,2}\rvert_{\theta=\nicefrac{\pi}{2}} = \phi_0\rvert_{\theta=\nicefrac{\pi}{2}} = 0\ .
\end{equation*}

\begin{figure*}[tp]
    \centering
    \includegraphics[width = 0.49\textwidth]{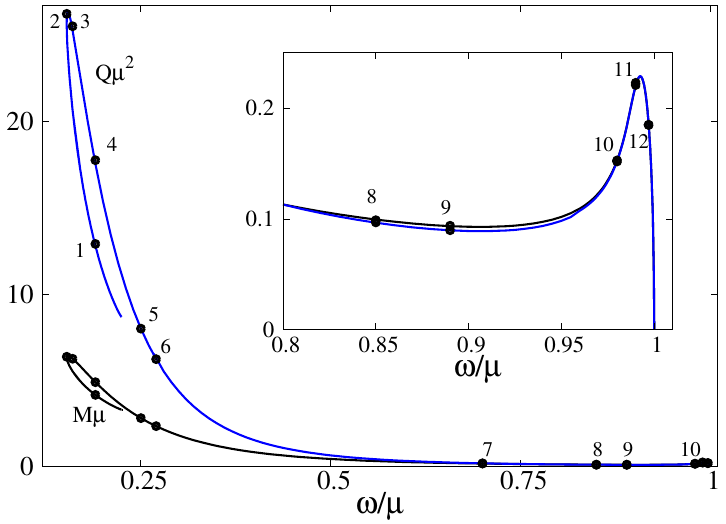} \hfill
     \includegraphics[width = 0.49\textwidth]{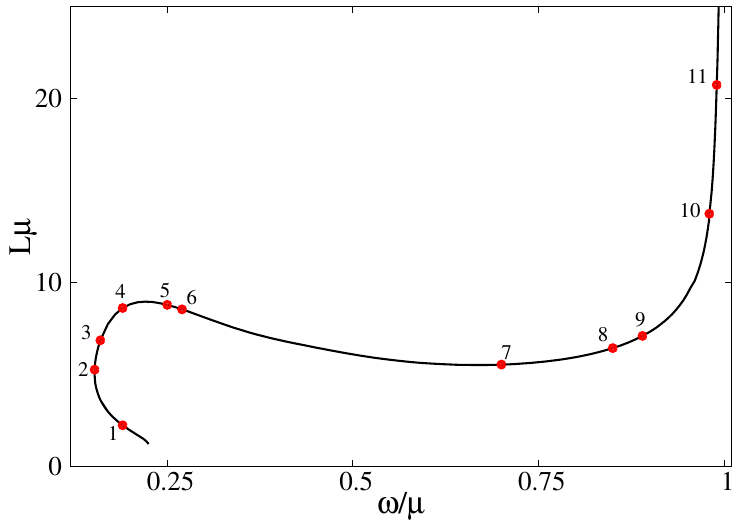}
    \caption{\emph{Dipolar Q-stars domain of existence}. (Left panel) ADM mass ($M$)/Noether charge ($Q$) $vs.$\ scalar field frequency $\left(\omega\right)$ diagram. The inset shows the behavior for the region close to the maximal frequency. Regions where the $M\mu>Q\mu^2$ are expected to be energetically unstable against fission. This occurs near the local minimum of the mass. (Right panel) Proper distance $L$  between the two components, or poles, of each star as a function of the scalar field frequency $\omega$. Notice the (non-monotonic) trend that the stars become closer when moving from the Newtonian to the relativistic branch.  The 12 highlighted points represent the solutions dynamically evolved below. Solution 12 is higher up on the right panel, outside the plot range.
    \label{fig:stable_DBS}}
\end{figure*}

The dipolar BSs  are static,  globally regular and
without an event horizon or conical singularities, and asymptotically flat. 
They possess two global charges.
The first one is the ADM mass $M$, 
which 
can be obtained from the respective Komar expression \cite{Wald:1984rg},
\begin{equation}
\label{komar}
{M} = \frac{1}{{4\pi }} \int_{\Sigma}
 R_{a b}n^a\xi^b \text{d}V \ ,~
\end{equation}
where $n^a$ is unit normal to $\Sigma$ 
and
$\text{d}V$ is the natural volume element on $\Sigma$. 
The ADM mass can also be read off from 
the asymptotic sub-leading behavior of the metric function $g_{tt}$
\begin{eqnarray}
\label{asym}
g_{tt} =-e^{2F_0}
=-1+\frac{2 M}{r}+  \dots \ .   
\end{eqnarray}   
There is also 
a conserved  Noether charge, computed from~(\ref{Q}) as
\begin{eqnarray}
\label{Q-int}
Q=4\pi \int_{0}^\infty \text{d}r \int_0^\pi\text{d}\theta  
~r^2\sin \theta ~e^{F_0+2F_1+F_2} \omega\phi^2 \ .
\end{eqnarray}

The energy and Noether charge densities of the different solutions are localized in two distinct components, named poles, located symmetrically on the $z$-axis and at $r= r_c$. The proper distance between these components is defined as,
\begin{equation}
    L = 2 \int_0^{r_c}\text{d}r\ e^{F_1(r, 0)}\ .
\end{equation}

\begin{figure*}[tpbh]
    \centering 
    \includegraphics[width=\linewidth]{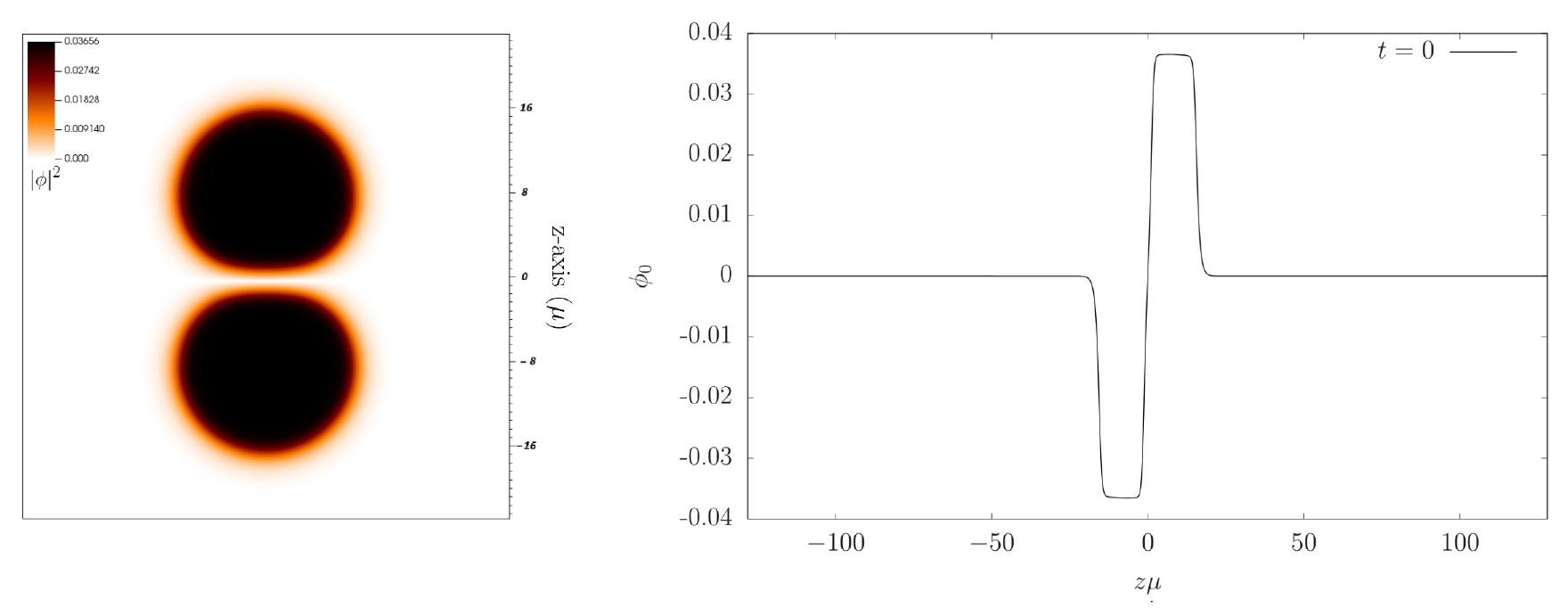}
    \includegraphics[width=\linewidth]{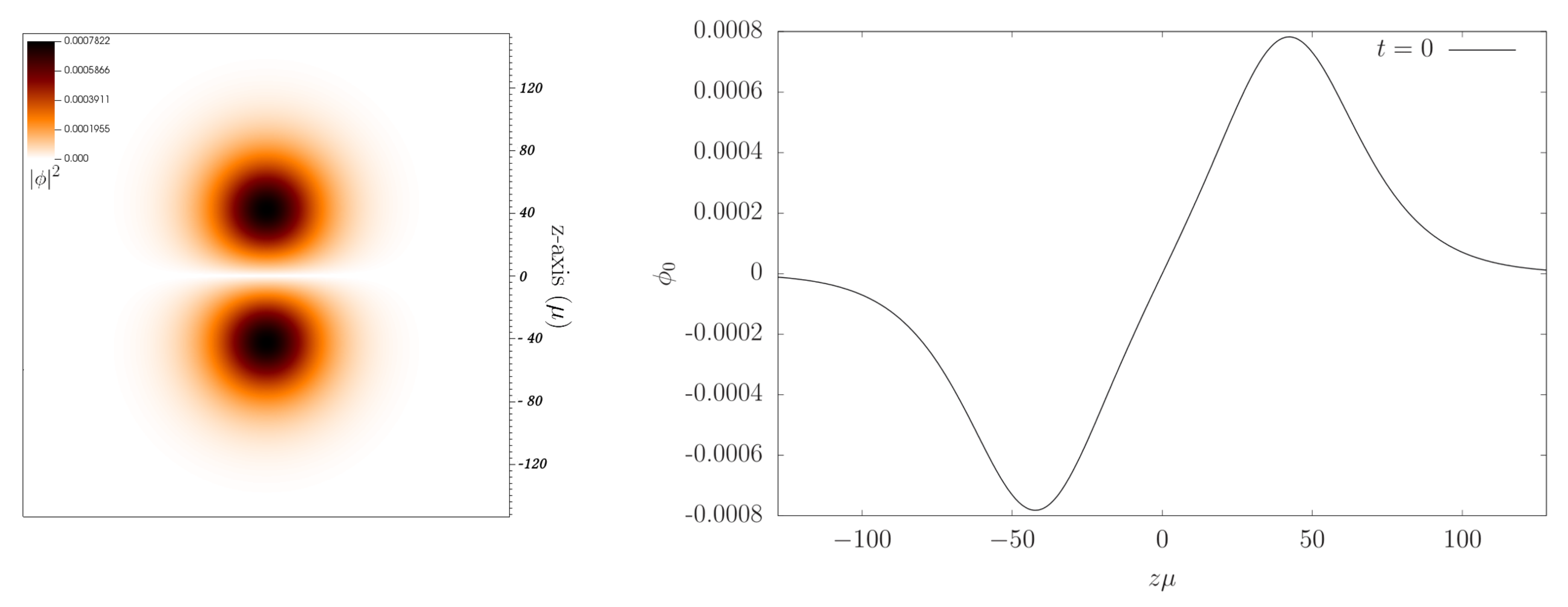}
    \caption{\emph{Two illustrative dipolar $Q$-stars (Solution 6 (top) and 12 (bottom) in Table~\ref{tab:DBSs_data})}.  (Left panels) 2-dimensional slice of the scalar field density $|\phi|^2$, on the $y=0$ plane. (Right panel) Scalar field amplitude $\phi_0$ along the $z$-axis. For the top solution, one observes the almost step-function profile, in contrast with mini-BSs, which have a less sharp spatial decay, which is approached here in the Newtonian branch, as illustrated by the bottom solution. These equilibrium solutions are the initial data for the dynamical evolutions in this paper. 
    \label{fig:1sol_ex}}
\end{figure*}

We now, and for the remainder of this paper, focus on dipolar $Q$-stars with $\sigma_0=0.05$, an even smaller value than those in Fig.~\ref{fig1}, making the stars even more compact. In Fig.~\ref{fig:stable_DBS} we give an overview of their domain of existence  (left panel), showing both the ADM mass and the Noether charge $vs.$  the scalar field frequency. One observes a similar structure as in other self-interacting BS models, including the monopolar $Q$-stars described above (\textit{e.g.}~\cite{Siemonsen:2020hcg,Guerra:2019srj,Delgado:2020udb,Cunha:2022gde}). 
Starting from the Newtonian limit, $\omega/\mu \to 1$, wherein BSs typically become very dilute and thus Newtonian, a first (local) maximum of the mass occurs at $\omega/\mu = 0.994$. The solutions between these two frequencies are the \textit{Newtonian branch}. Then the mass decreases to a local minimum at $\omega/\mu = 0.907$, whence it starts increasing again, reaching a global maximum at $\omega/\mu = 0.1522$. Within these two frequencies is the \textit{relativistic branch} and within the Newtonian and relativistic branch we have the \textit{middle} branch. The minimum frequency attained, which is below that delimiting the relativistic branch, occurs for $\omega/\mu = 0.1520$. The right panel of Fig.~\ref{fig:stable_DBS} shows how the proper distance between the two centers varies along the domain of existence.

\begin{table}[bth]
\centering
\caption{Selected dipolar $Q$-stars. \label{tab:DBSs_data}}
\begin{ruledtabular}
\begin{tabular}{@{}lccccc@{}}
Sol. & $\omega/\mu$ & $\phi_0\left(r=r_c\right)$ & $\mu M$ &  $\mu^2 Q$ & $\mu L$ \\ 
\hline
1 ($2$nd) & 0.1900  & 0.0442 & 4.149 & 12.910 &2.261  \\
2 & 0.1522  & 0.0376 & 6.363 & 26.272 & 5.277  \\
\hline
3 & 0.1600  & 0.0369 & 6.249 & 25.555 & 6.873 \\
4  & 0.1900  & 0.0365 & 4.889 & 17.774 & 8.624  \\
5 & 0.2500  & 0.0364 & 2.799 & 7.996 & 8.795  \\
6 & 0.2700  & 0.0365 & 2.333 & 6.226 & 8.550  \\
7 & 0.7000  & 0.0372 & 0.161 & 0.178 & 5.554  \\
8 & 0.8500  & 0.0319 & 0.099 & 0.097 & 6.449  \\
9 & 0.8900  & 0.0286 & 0.094 & 0.090 & 7.108  \\
\hline
10 & 0.9800  & 0.0097 & 0.153 & 0.152 & 13.741 \\
11 & 0.9900  & 0.0038 & 0.223 & 0.221 & 20.731 \\
\hline
12 & 0.9970 & 0.0008 & 0.185 & 0.185 & 42.463 
\end{tabular}
\end{ruledtabular}
\end{table}

\begin{figure}[htpb]
  	\centering
  	\includegraphics[width=0.5\textwidth]{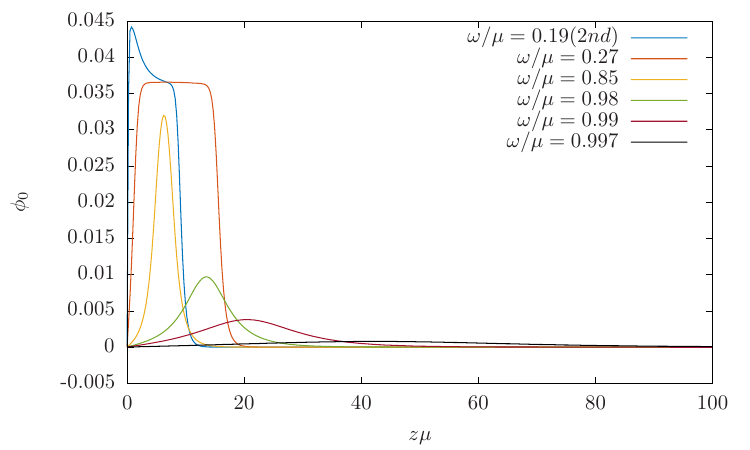}
  	\caption{\emph{$z$-profile of illustrative dipolar $Q$-stars}.  Real part of the scalar field along the $z$-axis, for solutions 1, 6, 8, 10, 11 and 12.
  	\label{fig:all_dqstar_phi1_t0} }
\end{figure}

Within the domain of existence we have selected 12  solutions, highlighted in Fig.~\ref{fig:stable_DBS}, with their physical properties detailed in Table~\ref{tab:DBSs_data}, that shall be considered in the dynamical evolutions below. The horizontal lines in the table separate the different branches defined above. To gain some insight into these solutions, Fig.~\ref{fig:1sol_ex} shows the morphology of two illustrative dipolar $Q$-stars. One can appreciate how compact the centers become in the relativistic branch, as opposed to the Newtonian branch. The scalar field profiles along the $z$-axis are also shown for seven of the chosen solutions in Fig.~\ref{fig:all_dqstar_phi1_t0}.
Of the 12 selected solutions, those with smaller frequencies, $\omega/\mu = \left\{ 0.1522,\ 0.16,\ 0.19,\, 0.25,\ 0.27 \right\}$, comprise highly compact and localised distributions of the scalar field. However, as we increase the scalar field frequency, the solutions become less compact and more dispersed across space, with each pole acquiring a similar shape to monopolar mini-BSs \cite{PhysRev.172.1331}. The latter trend is quite natural; as $\omega/\mu \rightarrow 1$, the scalar field amplitude decreases and eventually vanishes. In the scalar potential \eqref{eq:qball_U}, higher power terms of $\phi_0$  decrease faster with $\omega/\mu\rightarrow 1$, meaning that they are suppressed in the Newtonian limit, making $Q$-stars similar to mini-BSs in that limit. 
%

\section{Numerical framework}
\label{sec:frame}

To perform numerical evolutions we employ the standard 3+1
decomposition~\cite{Gourgoulhon:2007ue,Alcubierre08a}. The metric line element is written in the form
\begin{equation}
    \text{d}s^2 = - \alpha^2\text{d}t^2+\gamma_{ij} \left(\text{d}x^i+\beta^i\text{d}t \right) \left(\text{d}x^j+\beta^j\text{d}t \right)\ ,
\label{eq:line_elem}
\end{equation}
where $\alpha$ is the lapse function, $\beta^i$ is the shift vector, and $\gamma_{ij}$ is the induced metric in each spatial foliation. We also introduce the extrinsic curvature
\begin{equation}
    K_{ij} =\ -\frac{1}{2 \alpha} \left(\partial_t - \mathcal{L}_\beta  \right) \gamma_{ij}\ ,
    \label{eq:extrinsic_curv}
\end{equation}
and, analogously, the ``canonical momentum'' of the complex scalar field $\phi$
\begin{equation}
    K_{\phi} =\ -\frac{1}{2 \alpha} \left(\partial_t - \mathcal{L}_{\beta}  \right) \phi\ ,
    \label{eq:canonical_mom}
\end{equation}
where $\mathcal{L}$ is the Lie derivative. In this form, the full EKG system of equations reads
\begin{align}
    \partial_{t} \gamma_{i j}& = -2 \alpha K_{i j}+\mathcal{L}_{\beta} \gamma_{i j}\ , \\
    \partial_{t} K_{i j} & = -D_{i} \partial_{j} \alpha+\alpha\left(R_{i j}-2 K_{i k} K_{j}^{k}+K K_{i j}\right) \nonumber \\
    & \quad + \mathcal{L}_{\beta} K_{i j} + 4 \pi \alpha \left[ \left(S - \rho\right) \gamma_{i j} - 2 S_{i j} \right]\ , \\
    \partial_{t} \phi &= -2 \alpha K_{\phi}+\mathcal{L}_{\beta} \phi\ , \\
    \partial_{t} K_{\phi} &= \alpha \Big[K K_{\phi}-\frac{1}{2} \gamma^{i j} D_{i} \partial_{j} \phi \nonumber \\
    & \quad + \frac{1}{2} \mu^2 \phi \left( 1 - 8 \frac{|\phi|^2}{\sigma^2_0} + 12 \frac{|\phi|^4}{\sigma^4_0} \right) \Big]\nonumber \\
    & \quad - \frac{1}{2} \gamma^{i j} \partial_{i} \alpha \partial_{j} \phi+\mathcal{L}_{\beta} K_{\phi}\ .
\end{align}
This system of equations is subjected to the set of constraints 
\begin{align}
    \mathcal{H} &\equiv R + K^2 - K_{ij} K^{ij} = 16 \pi \rho\ , \label{eq:ham_constraint}\\
    \mathcal{M}_{i} &\equiv D_i K - D^j K_{ij} = - 8 \pi j_i\ , \label{eq:mom_constraint}
\end{align}
where $D_{i}$ denotes the covariant derivative with respect to the 3-metric $\gamma_{ij}$. The source terms are given by
\begin{align*}
    \rho & \equiv T_{ab} n^a n^b\ , \\
    j_i  & \equiv - \gamma_{i}^a T_{ab} n^b\ , \\
    S_{ij} & \equiv \gamma_{\ i}^a \gamma_{\ j}^b T_{ab}\ , \\
    S & \equiv \gamma^{i j} S_{ij}\ ,
\end{align*}
where $\rho$, $j_i$, $S_{ij}$ and $S$ denote the energy density, momentum density, stress, and the trace of the stress as observed by a normal observer (moving along the normal vector $n^a$), respectively.

For numerical evolutions, the equations above are rewritten in the strongly hyperbolic BSSN (Baumgarte-Shapiro-Shibata-Nakamura) scheme~\cite{PhysRevD.59.024007, PhysRevD.52.5428},
and numerically evolved using the \textsc{EinsteinToolkit} (ET) \cite{zilhao2013introduction, EinsteinToolkit:2021_05} infrastructure.  Our numerical implementation uses the BSSN evolution system as detailed in Ref.~\cite{Cunha:2017wao}. The spacetime metric and scalar field variables are evolved in time using the \textsc{LeanBSSNMoL} and \textsc{ScalarEvolve} \textsc{Cactus} \emph{thorns} \cite{Canuda_zenodo_3565474}.
We use the \textsc{Carpet}~\cite{Schnetter:2003rb} library for mesh refinement capabilities and \textsc{AHFinderDirect}~\cite{Thornburg:1995cp,Thornburg:2003sf}
for finding apparent horizons.

\section{Results}
\label{sec:Dipolar Boson Star solutions}

With the framework outlined in the previous section, we evolve the dipolar $Q$-stars using the equilibrium solutions described in Sec.~\ref{sec:backg} as initial data. The numerical evolutions are performed in units where $ \mu \sigma_0 \sqrt{8 \pi} = 1$. For the solutions considered herein, we have fixed $ \sigma_0 = 0.05$.
\begin{figure}[htbp]
    \centering
    \includegraphics[width=0.475\textwidth]{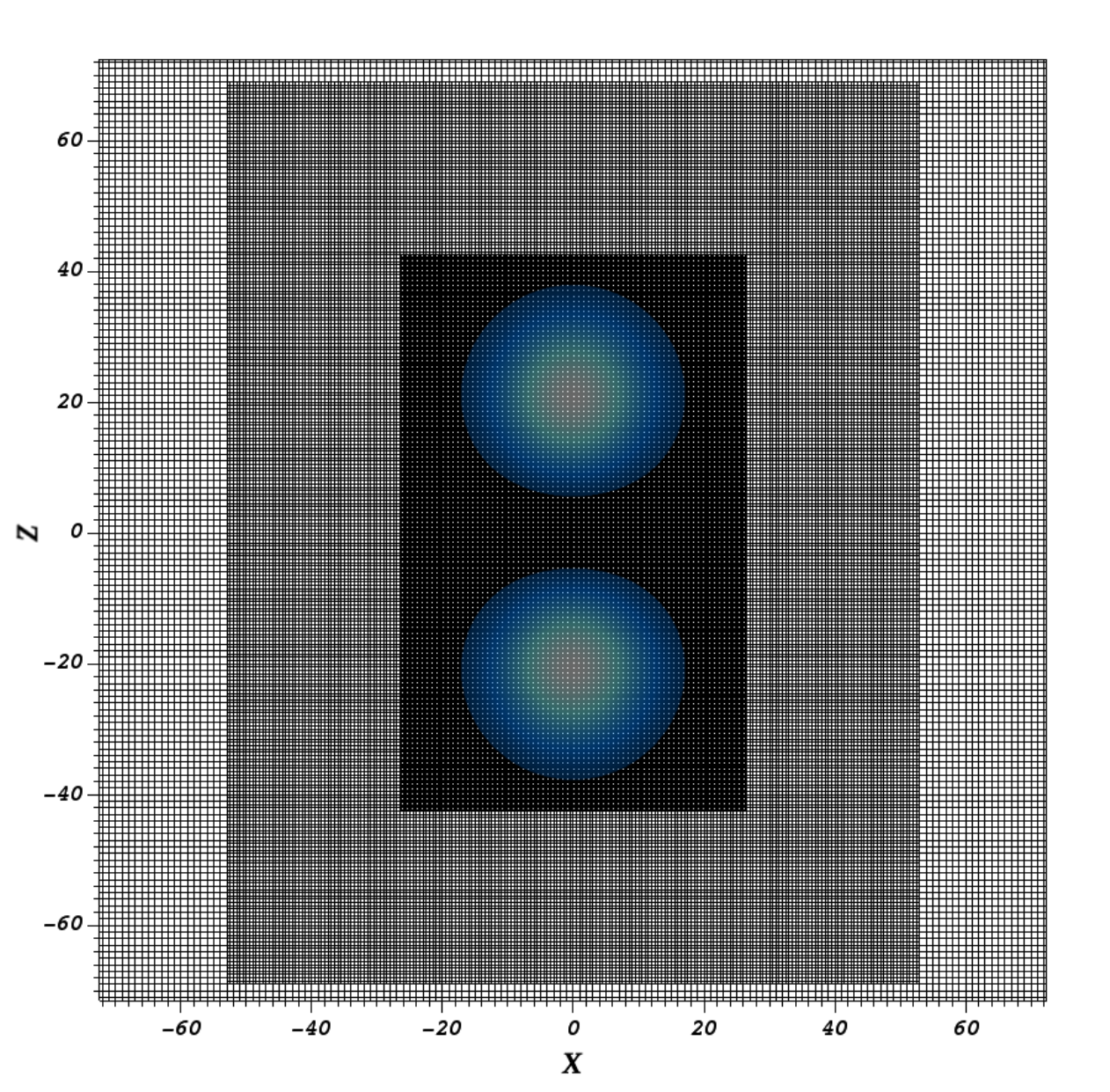}
    \caption[\emph{Grid for dipolar $Q$-stars evolutions}. Grid configuration for the evolution of dipolar Q-stars]{\emph{Grid for dipolar $Q$-stars evolutions}. The grid is zoomed out so that the three refinement levels, with resolutions of, from the innermost to the outermost, $h\mu = \left\{ 0.25,\ 0.50,\ 1.00 \right\}$, can be clearly viewed. Here $X\equiv x \mu$ and $Z \equiv z \mu$. 
    \label{fig:dipole_grid}}
\end{figure}
All solutions were evolved numerically in a grid with three refinement levels -- see Fig.~\ref{fig:dipole_grid} for a typical configuration.
The grid has a rectangular shape on the two innermost levels and an overall size of $x\mu,y\mu \in \left[ 0,\ +128 \right]$ and $z\mu\in \left[ -128,\ +128 \right]$. We impose symmetry on the $x$ and $y$-axis given that the solutions are axisymmetric and the dipole is oriented along the $z$-axis. For all but solution $12$, the innermost level has a grid spacing of $h\mu=0.25$.
For solution 12, given its large radius and our numerical limitations, we increased the grid spacing in the innermost level 
to $h\mu=0.5$. 

\begin{figure}[htbp]
  	\centering
  	\includegraphics[width=0.45\textwidth]{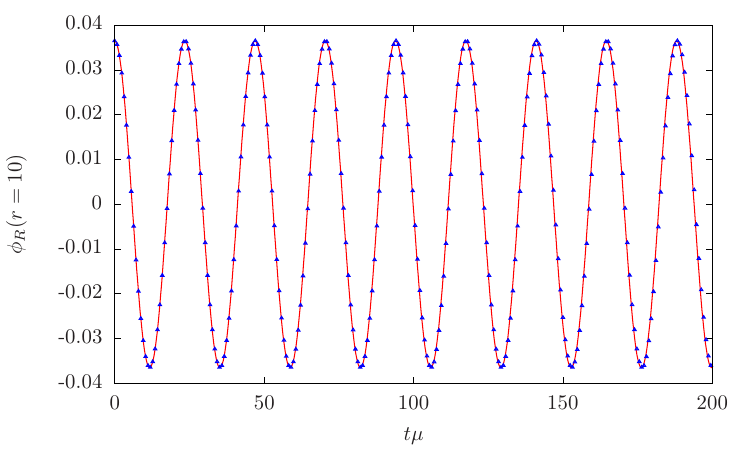}
  	\caption{\emph{Dipolar $Q$-star solution 6 with $\omega / \mu = 0.27$}. Evolution of the real part of $\phi_0$ at $ r_{\textrm{ex}} \mu = 10$. The analytical expected value -- $0.0365 \cos{\left(0.27t\right)}$ -- is illustrated as the red line while the numerical evolution is shown by the blue points. 
  	\label{fig:dqstar_0.25_freq}}
\end{figure}

In order to verify the agreement between the numerical and the analytical evolutions ($i.e.$ for a star in equilibrium, the phase evolution is dictated by Eq.~\eqref{eq:scalar_field_axia}), we have compared the numerical output of the oscillation of the real part of the scalar field, $\phi_R$, with its analytical counterpart -- $\phi_0(r_\mathrm{ex}) \cos(\omega t)$, where $r_{\rm ex}$ is the distance from the origin (along the axis) at which the numerical output is extracted. 
We have observed complete agreement between the numerical and the analytical data for all 12 solutions. We illustrate this analysis for solution 6 in Fig.~\ref{fig:dqstar_0.25_freq}.

\subsection{Colapsing dipoles beyond the relativistic branch}

\begin{figure}[htbp]
  	\includegraphics[width=0.49\textwidth]{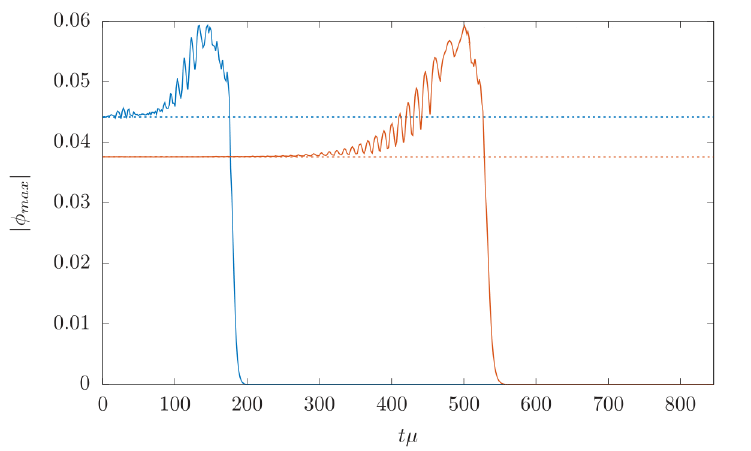} \hfill
    \includegraphics[width=0.49\textwidth]{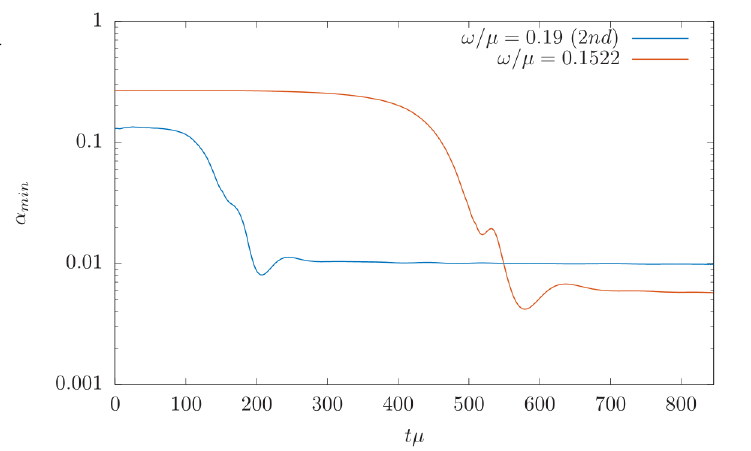}
  	\caption{\emph{Collapsing dipolar $Q$-stars, Solutions 1 and 2.} (Top panel) Time evolution of the maximum value of the scalar field. It remains approximately constant up until the instant of collapse. The dashed lines represent the scalar field at the center of the poles (the region of maximum density) of each solution at $t=0$. The collapse to a more compact object is signalled by the slight increase in the scalar field density followed by its drop to zero, hinting at scalar matter crossing the BH horizon.
    (Bottom panel) Time evolution of the minimum value of lapse function. It remains approximately constant up until the instant of collapse, after which its value decreases exponentially to zero.
  	\label{fig:lapse_scalar_unst_dbs}}
\end{figure}

Let us start with the most compact dipoles, in the sense of the right panel of Fig.~\ref{fig:stable_DBS}. We observe that solutions 1 and 2, placed to the left and on the absolute maximum of the mass, respectively -- see Fig.~\ref{fig:stable_DBS} (left panel) -- 
undergo gravitational collapse shortly after the beginning of the simulation. 
This can be seen in Fig.~\ref{fig:lapse_scalar_unst_dbs}, where both the maximum of the scalar field and the minimum of the lapse function are plotted as functions of time.
%
Typically, the ``collapse of the lapse'' (where the lapse function, responsible for quantifying the proper time between each spacelike slice, falls exponentially to zero) signals the formation of an apparent horizon.
In Fig.~\ref{fig:lapse_scalar_unst_dbs} we can indeed see that when the lapse function drops abruptly, so does the maximum of the scalar field, indicating that matter is being swallowed by the newly formed BH.
This result is in accordance with what would occur in the corresponding region of the domain of existence for the fundamental monopolar $Q$-stars of the model. Solution 1 is beyond the relativistic branch and solution 2 sits on its edge. The instabilities of solutions beyond the relativistic branch have been observed in other models of self-interacting bosonic stars and appear to be a general feature -- see \textit{e.g.}~\cite{Siemonsen:2020hcg}.

\begin{figure*}[tbp]
  	\includegraphics[width=0.49\textwidth]{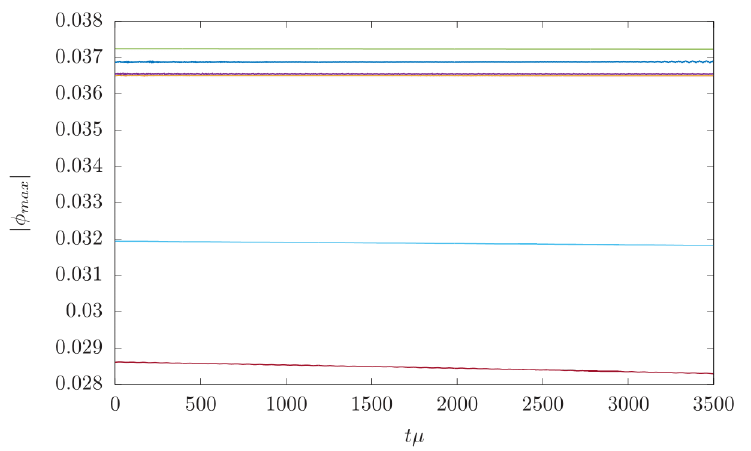} \hfill
  	\includegraphics[width=0.49\textwidth]{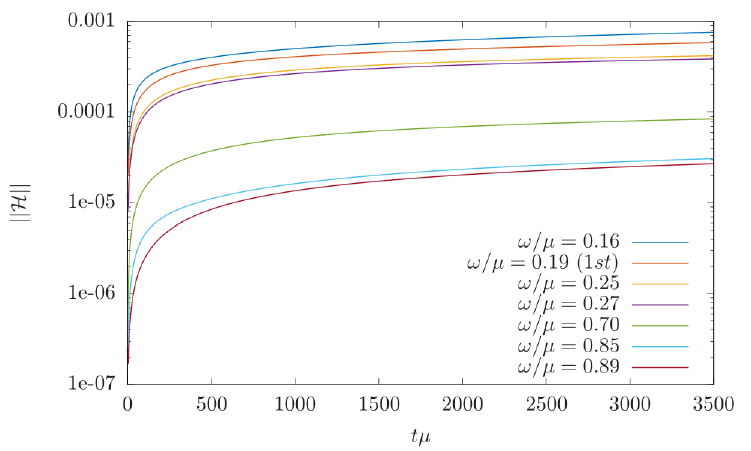}
  	\caption{\emph{Dynamically robust dipolar Q-stars sitting in the relativistic stable branch, solutions 3--9 in Table~\ref{tab:DBSs_data}.} (Left panel) Time evolution of the maximum value of the scalar field's density. (Right panel) Time evolution of the $L_2$ norm of the Hamiltonian constraint.
  	\label{fig:scalar_H_stbl_dbs}}
\end{figure*}

\begin{figure*}
    \centering
   \includegraphics[width=1\textwidth]{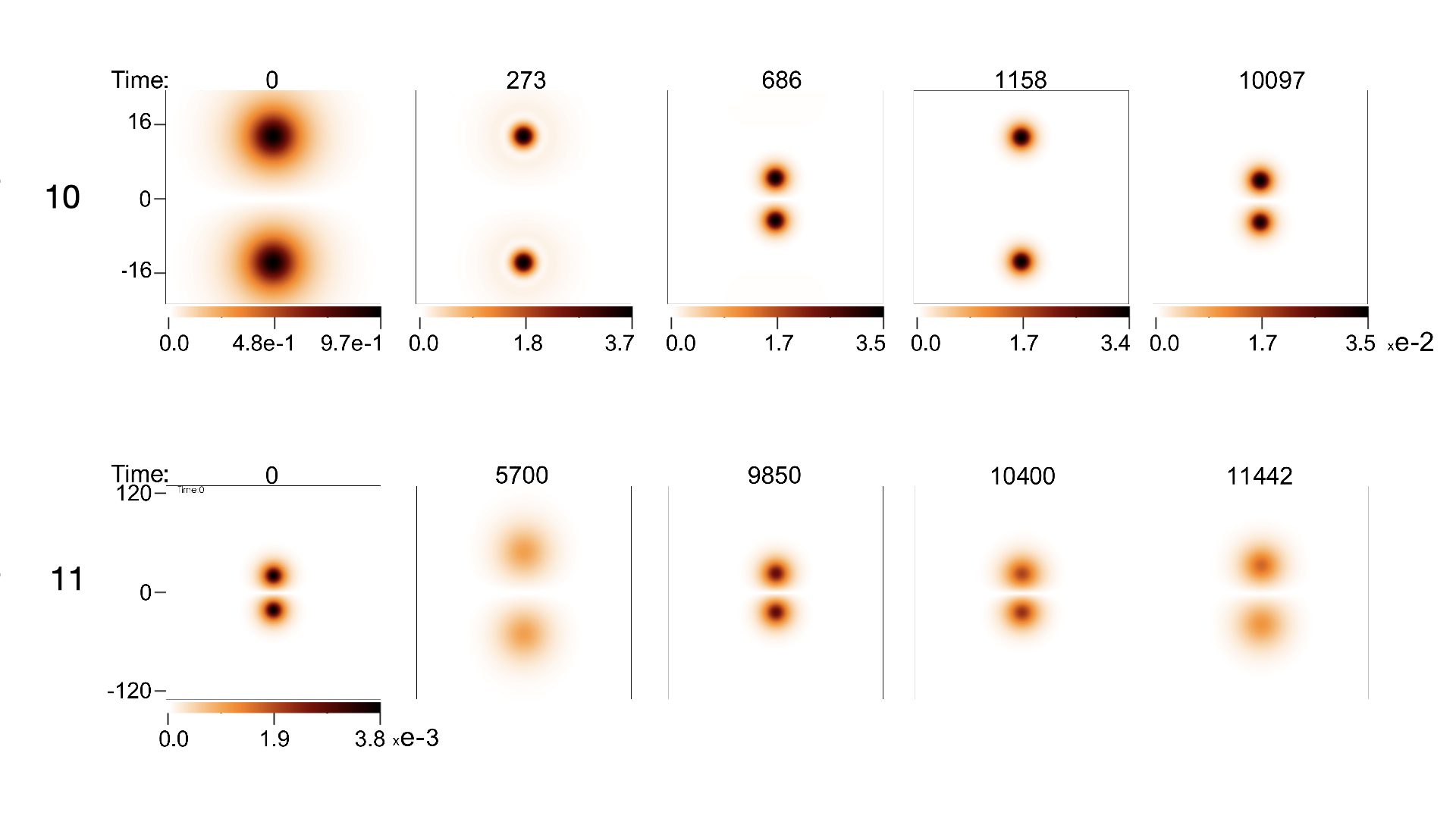}
    \caption{\emph{Evolution of unstable dipolar Q-stars}. Snapshots of the scalar field density on the $y=0$ plane, for solution 10 (top row) and 11 (bottom row).  The horizontal axis of each panel has the same spatial scale as its vertical axis and the color scalar in the bottom row is the same for all snapshots.
    \label{fig:snapshots_dipolarqstars}}
\end{figure*}

\subsection{Robust dipoles in relativistic and Newtonian branches}

Next, we consider simultaneously the solutions both in the relativistic branch (3--9) and in the Newtonian branch (12). These solutions  showed no evidence of unstable behavior during their simulation time, a minimum of $t\mu \sim 3500$. To illustrate this lack of change, we present in Fig.~\ref{fig:scalar_H_stbl_dbs} the time evolution of both the maximum value of the scalar field and the $L_2$ norm of the violation of the Hamiltonian constraint, respectively, for all seven solutions in the relativistic branch. 
As can be seen, the scalar field density of each star remains approximately constant during the simulation time, without dramatic changes. Note that the simulation time is much larger than the one where collapse is observed for solutions 1--2 -- \textit{cf.}~Fig.~\ref{fig:lapse_scalar_unst_dbs}.

\subsection{Unstable dipoles in middle branch}
%
Now we consider the two illustrative solutions in the middle branch, within the relativistic and Newtonian branches (10--11).
These solutions present an unstable behavior, but with two qualitatively different evolutions, that were followed up to $t \mu \sim 10000$. These are exhibited in Fig.~\ref{fig:snapshots_dipolarqstars}, where the qualitative distinction can be appreciated.

\subsubsection*{Development  of the instability}
\begin{figure}[tp]
    \includegraphics[width=0.49\textwidth]{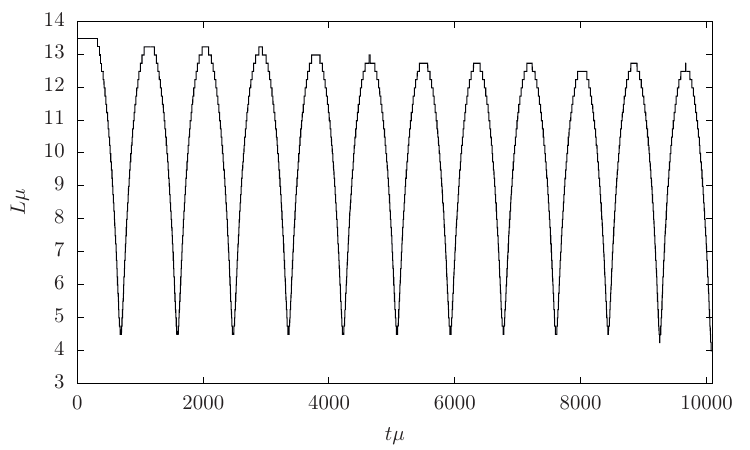}
    \includegraphics[width=0.49\textwidth]{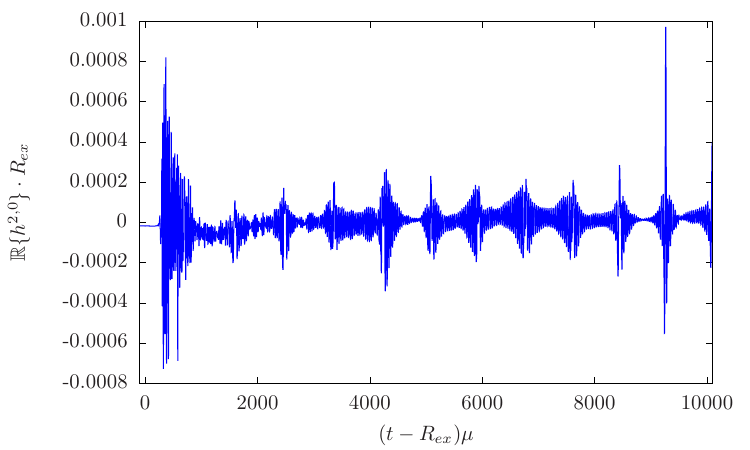}
  	\caption{\emph{Evolution of solution 10:} (Top panel) The distance to the origin, along the $z$-axis, of one of the poles of solution 10, as a function of time. The maximum distance reached after each bounce decreases with time due to energy loss via gravitational radiation. For a simulation time of $t \mu \sim 12500$, solution 10 performed 12 collisions, oscillating roughly within a distance $\left[4.5,\ 13.5 \right]$ from the origin. (Bottom panel) The real part of the $(\ell,m)=(2,0)$ mode of the Newman-Penrose scalar $\Psi_4$, describing the gravitational wave emission of solution 10 extracted at $R_{\textrm{ex}}\mu = 100$, as a function of time.
  	\label{fig:centroid_dqstar_10_11}}
\end{figure}

%
Consider first solution 10. It exhibits a noticeable change at a fairly short time scale of $t\mu\sim 100$. The two individual centres become more compact, accompanied by the ejection of part of the scalar field. The corresponding newly formed dipole is, however, off balance, resulting in a \textit{dynamical} dipole. The scalar field repulsion between the poles ceases to be able to hold the gravitational pull after the initial readjustment, and the poles begin to move towards each other. Eventually, these collide inelastically and rebound back to close (but not quite) their initial positions, which we define as the \textit{rebound distance}, whence they fall back into each other again, repeating this process a number of times over the duration of the simulation, with the rebound distance trending towards a decrease after each collision --  Fig.~\ref{fig:centroid_dqstar_10_11} (top panel). This decrease can be explained by the loss of linear momentum via gravitational waves emission, as shown in Fig.~\ref{fig:centroid_dqstar_10_11} (bottom panel).
For solution 10, this process results in $12$ collisions for a simulation time of $t \mu \sim 12500$, but a larger number of collisions was observed in the simulations of other dipolar $Q$-star solutions near solution 10 (not shown here).
The overall evolution after the dipole becomes dynamical, with sequences of collisions, is reminiscent of the head-on collisions of (monopolar) BSs in this model~\cite{PhysRevD.95.124005}.

Solution 11 presents a somewhat opposite behavior to that of solution 10. The key difference is that the individual centers become \textit{less} compact: there is a clear, but slow, expansion of the scalar field distribution of each pole that is halted at $t\mu \sim 5700$. The solution then contracts again, returning to a configuration similar to that of its initial data before expanding again -- Fig.~\ref{fig:snapshots_dipolarqstars} (lower panel).

\subsubsection*{Endpoint of the instability}



The behaviour indicated for solutions 10 and 11 suggests a migration to other solutions with a different scalar field frequency $\omega/\mu$ and scalar field amplitude $\phi_0$.
One way to probe this migration and attempt to unveil the endpoint is by analysing the evolution of these quantities, an analysis we now describe. 

\begin{figure}[t]
    \centering
  	\includegraphics[width=0.5\textwidth]{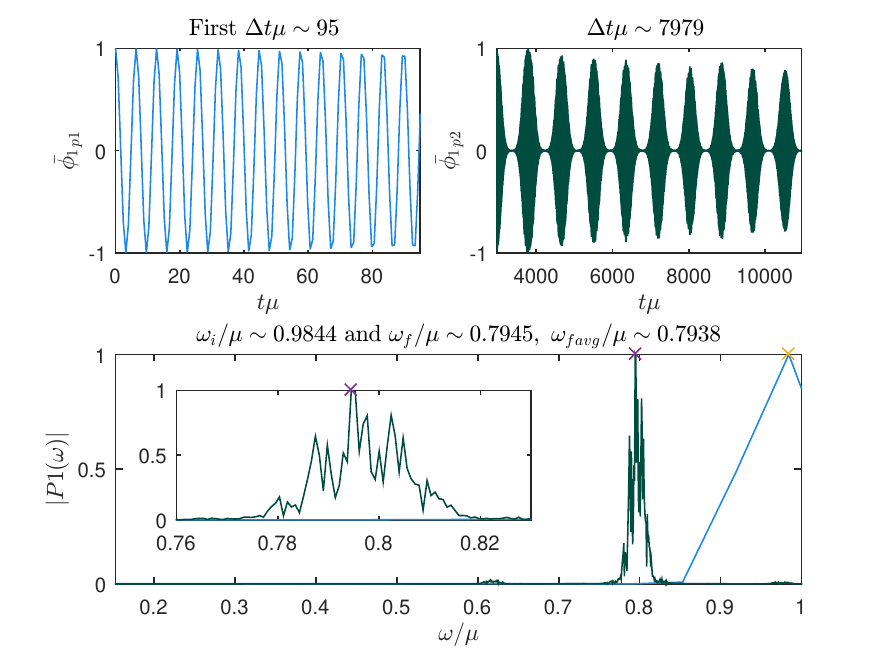}
  	\caption{\emph{Tracking the evolution endpoint for solution 10}. (Top left panel) The normalised real component of the scalar field as a function of time for the first $\Delta t \mu \sim 95$ (initial period)  of the evolution. (Top right panel) The normalised real component of the scalar field as a function of time for the last $\Delta t \mu \sim 7979$ (final period) of the evolution. (Bottom panel) A clear transition is seen from the initial to the final period, from a higher to a lower and more dispersed value of the scalar field frequency. Their values are $\omega/\mu = 0.9844$ for the initial period and $\omega/\mu \sim 0.7945$ for the final period.}
  	\label{fig:freq_fft_dqstar_0.98}
\end{figure}

\begin{figure}[t]
    \centering
  	\includegraphics[width=0.5\textwidth]{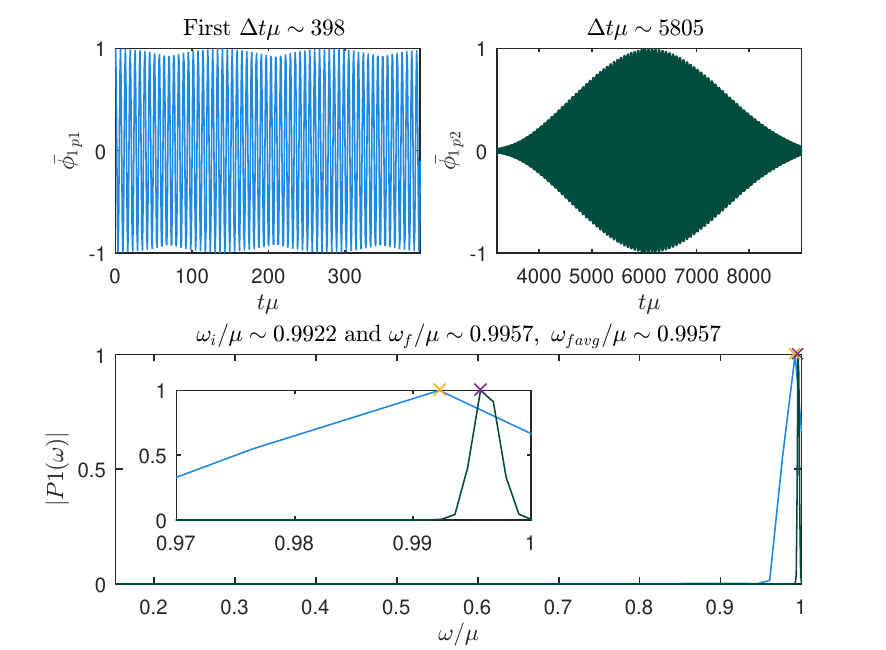}
  	\caption{\emph{Tracking the evolution endpoint for solution 11}. (Top left panel) The normalised real component of the scalar field as a function of time for the first $\Delta t \mu \sim 398$ (initial period)  of the evolution. (Top right panel) The normalised real component of the scalar field as a function of time for the last $\Delta t \mu \sim 5805$ (final period)  of the evolution. (Bottom panel) A clear transition is seen from the initial to the final period, from a lower to a higher and less dispersed value of the scalar field frequency. Their values are $\omega/\mu = 0.9922$ for the initial period and $\omega/\mu \sim 0.9957$ for the final period.}
  	\label{fig:freq_fft_dqstar_0.99}
\end{figure}

\begin{figure*}[tp]
    \centering
    \includegraphics[width = 0.8\textwidth]{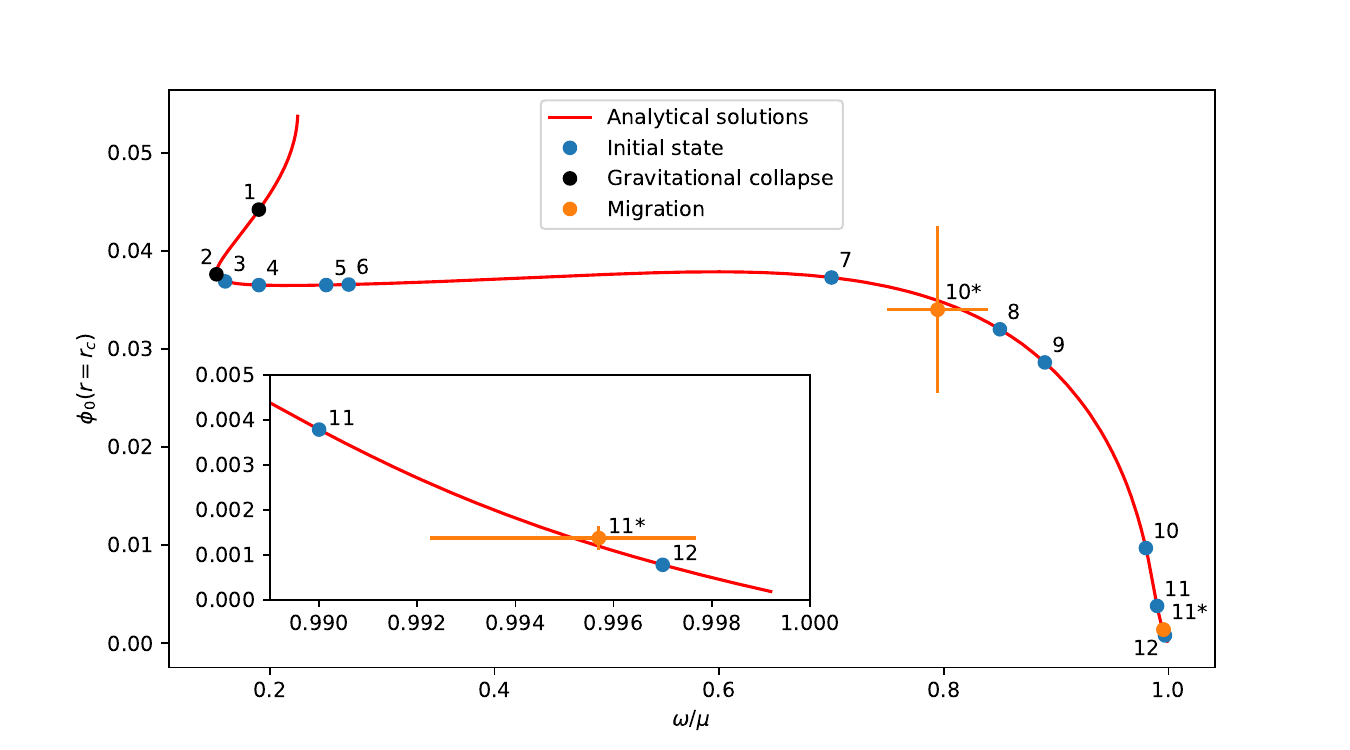}
    \caption{\emph{Domain of existence of dipolar $Q$-stars}. Maximum scalar field amplitude $(\phi_0)$ \textit{vs.}  scalar field frequency $\left(\omega / \mu\right)$ diagram. The solutions presented correspond to the final states of all 12 solutions. Migrating solutions 10 and 11 are depicted both in their initial state (orange circles) and in their provisional final state (blue circles). The inset shows the behaviour for the region close to the maximal frequency. The error bars for solutions 10 and 11 illustrate the oscillations in the scalar field amplitude due to their still evolving state, and the standard deviation in the frequency domain due to the detection of a range of frequencies when applying a Fast Fourier transform to the signal.}
    \label{fig:DBS_migrt}
\end{figure*}

We begin by analysing the scalar field frequency and its amplitude. A Fourier analysis \cite{FFTW05} is performed on the real part of the scalar field, as is shown for the two solutions in Figs.~\ref{fig:freq_fft_dqstar_0.98} and  \ref{fig:freq_fft_dqstar_0.99}.
Within the range of validity, $\omega/\mu \in \left[0.1522,\ 1 \right]$, we find that solution 11 acquires higher oscillation frequencies, within the range of the Newtonian branch, wherein solution 12 is located, which was seen as dynamically robust in our analysis.  In contrast, despite displaying several peaks for the oscillation frequency, which might indicate different frequencies acquired during its migration, solution 10 acquires an average frequency well below its initial one and within the range of the relativistic branch, close to solution number 7, proven stable. An overview of this state of affairs is exhibited in Fig.~\ref{fig:DBS_migrt}, where the frequency $\omega$ of each solution is plotted against its scalar field amplitude $\phi_0$ at the center of one of its poles, at $t=0$ (``Initial state'') for all solutions, and at the end of the numerical evolution (``Migration'') for solutions number 10 and 11.  

This evidence points towards the migration of unstable solutions to either the relativistic or the Newtonian branches, wherein solutions show a higher degree of dynamical robustness. The initial and final states of every solution in our analysis are summarized in Table \ref{tab:DBSs_final}.
\begin{table*}[tb]
\centering
\caption{Fate of the dipolar $Q$-star solutions with different values of the scalar field oscillation frequency $\omega$.\label{tab:DBSs_final}}
\begin{ruledtabular}
\begin{tabular}{lcccccc}
Solutions & Initial branch & Initial frequency & Simulation duration &
Dynamical status & Final frequency & Final branch \\ 
\hline
1  & Beyond Relativistic & 0.1900 & 400   & Gravitational collapse  & --   & --          \\
2  & Boundary Relativistic & 0.1522 & 600   & Gravitational collapse  & --   & --          \\
3  & Relativistic & 0.1600 & 5300  & Stable    & 0.1600 & Relativistic \\
4  & Relativistic & 0.1900 & 5400  & Stable    & 0.1900 & Relativistic \\
5  & Relativistic & 0.2500 & 6500  & Stable    & 0.2500 & Relativistic \\
6  & Relativistic & 0.2700 & 5800  & Stable    & 0.2700 & Relativistic \\
7  & Relativistic & 0.7000 & 4500  & Stable    & 0.7000 & Relativistic \\
8  & Relativistic       & 0.8500 & 9000  & Stable    & 0.8500 & Relativistic       \\
9  & Relativistic       & 0.8900 & 7000  & Stable  & 0.8900 & Relativistic          \\
10 & Middle       & 0.9800 & 11000 & Migration & 0.7983    & Relativistic\footnote{At the boundary between the relativistic branch $\omega/\mu < 0.907$ and the middle branch $0.907 < \omega/\mu < 0.9924$.}        \\
11 & Middle       & 0.9900 & 11400 & Migration & 0.9957 & Newtonian    \\
12 & Newtonian    & 0.9970 & 4500  & Stable    & 0.9970 & Newtonian
\end{tabular}
\end{ruledtabular}
\end{table*}

\section{Conclusions}
\label{sec:concl}
In this paper, we have constructed dipolar BSs with sextic (+ quartic) self-interactions, according to the potential~\eqref{eq:qball_U}, named dipolar $Q$-stars, and analyzed their dynamics via fully non-linear numerical relativity simulations.

Our motivation was two-fold. Firstly, dipolar BSs constructed in the scalar model without self-interactions have exhibited an instability that develops in a timescale $t\mu\lesssim 2000$ for the models studied in~\cite{PhysRevLett.126.241105}. Secondly, since such dipolar stars can be seen as a type of excited state (as $p$-orbitals in hydrogen) -- with higher mass than the corresponding spherical stars with the same frequency (\textit{cf.}~Fig~\ref{fig1}); given the potential scalar self-interactions have already shown to mitigate dynamical instabilities of excited models, namely with rotation~\cite{Sanchis-Gual:2019ljs} and radially excited~\cite{Sanchis-Gual:2021phr,Brito:2023fwr} (which are akin to $Ns$-orbitals with $N>1$ in hydrogen), it becomes interesting to probe the impact of self-interactions on the stability of the dipolar $Q$-stars.

Our construction of the equilibrium solutions, presented in Section~\ref{sec:backg}, showed a domain of existence akin to that of the monopolar stars in the same model -- see Fig.~\ref{fig1} (bottom panel). In the case of the fundamental monopolar stars, such domain of existence includes a Newtonian and a relativistic branch wherein (spherical) $Q$-stars are stable, separated by a middle branch wherein stars are unstable -- see Fig.~\ref{fig2}. Moreover, beyond (to lower frequencies) the relativistic stable branch, $Q$-stars become too compact and unstable, forming BHs.

Here, we have studied the dynamical robustness of dipolar $Q$-stars by presenting a sample of evolutions of 12 illustrative solutions, covering different branches -- see Fig.~\ref{fig:stable_DBS} and Table~\ref{tab:DBSs_data}. 
Our evolutions provide evidence of similar dynamical properties for the dipolar $Q$-stars as the ones observed from the perturbative analysis of their spherical $Q$-star counterparts, namely: ($i$) in the Newtonian and relativistic branches they are dynamically robust over time scales longer than those for which dipolar stars without self-interactions are seen to decay, which were mentioned above; ($ii$) in the middle branch the dipolar $Q$-stars appear to migrate to either the Newtonian or the relativistic branch; ($iii$) beyond the relativistic branch, they decay to BHs.

There are, however, some caveats, in particular concerning the unstable states in the middle branch, that we should comment on. 
Solutions 10 and 11 showed evidence of a possible migration mechanism that allows the migration to different, dynamically more robust, solutions. However, at the end of our simulations, the solutions remain dynamical. We have no clear evidence for any dramatic effect altering the evolution, but we cannot rule it out either. Moreover, for very long time evolutions it becomes challenging to disentangle physical effects from numerical artifacts, sourced by accumulated errors. As such, the final state still requires further investigation. 

It may be that all these dipolar $Q$-stars are mere transient states. In fact, solutions 8 and 9, for which no instability was seen, have a slight energy excess, suggesting an energetic instability. This is reminiscent of an energetic instability seen for rotating BSs with self-interactions~\cite{Siemonsen:2020hcg}, occurring in the putative relativistic stable branch, but where fragmentation into a binary of non-rotating stars becomes dynamically favourable. Still, in our simulations, this possible energetic instability did not manifest itself and had no impact on the dynamics in the timescales probed. 

On the other hand, what our analysis could establish is that solutions within the relativistic stable branch and those in the Newtonian branch present stability time scales well above those of dipolar BSs without self-interactions. Moreover, the unstable solutions in the middle branch \textit{remain} dipolar, but dynamical ones, when readjusting their distance and compactness towards a more favourable configuration. From another perspective, these become interesting head-on collisions of $Q$-stars, with initial data obeying all constraints, in fact resembling previously studied head-on collisions in this model~\cite{PhysRevD.95.124005}.

It would be interesting to extend this analysis to spinning dipolar mini-BSs and $Q$-stars, in particular, due to their capacity to harbour one~\cite{Kunz:2019bhm}, or two~\cite{Herdeiro:2023roz} BHs in equilibrium with this scalar environment.
\begin{acknowledgments}
We would like to thank N. Sanchis-Gual for many useful discussions. N. M. Santos is supported by the FCT grant SFRH/BD/143407/2019. We acknowledge financial support by the Center for Research and Development in Mathematics and Applications (CIDMA) through the Portuguese Foundation for Science and Technology (FCT -- Fundação para a Ciência e a Tecnologia) -- references UIDB/04106/2020 and UIDP/04106/2020 -- as well as FCT projects 2022.00721.CEECIND, CERN/FIS-PAR/0027/2019, PTDC/FIS-AST/3041/2020, CERN/FIS-PAR/0024/2021, PTDC/MAT-APL/30043/2017 and 2022.04560.PTDC.
This work has further been supported by the European Horizon Europe staff exchange (SE) programme HORIZON-MSCA-2021-SE-01 Grant No.\ NewFunFiCO-101086251.
All simulations were performed with the Minho Advanced Computer Center (MACC) and Infrastrutura Nacional de Computação Distribuída (INCD) Cirrus-B clusters at the University of Minho, and the Baltasar clusters at IST.
\end{acknowledgments}

\bibliography{references}

\end{document}